\documentclass[aps,twocolumn,superscriptaddress]{revtex4-1}

\usepackage{graphicx}
\usepackage{subfigure}
\usepackage{amsmath}
\usepackage{mathrsfs}
\usepackage{amsthm}
\usepackage{amssymb}
\usepackage{hyperref}
\usepackage{xcolor}

\hypersetup{colorlinks=true, linkcolor=black, citecolor=black, urlcolor=blue}
\allowdisplaybreaks

\begin{document}

\newcommand{\ketbra}[2]{|#1\rangle\!\langle#2|}
\newcommand{\bra}[1]    {\langle #1|}
\newcommand{\ket}[1]    {| #1 \rangle}
\newcommand{\braket}[2]    {\langle #1 | #2 \rangle}
\newcommand{\braketbig}[2]    {\big\langle #1 \big| #2 \big\rangle}
\newcommand{\braketbigg}[2]    {\bigg\langle #1 \bigg| #2 \bigg\rangle}
\newcommand{\tr}[1]    {{\rm Tr}\left[ #1 \right]}
\newcommand{\av}[1]    {\langle{#1}\rangle}
\newcommand{\avbig}[1]    {\big\langle{#1}\big\rangle}
\newcommand{\avbigg}[1]    {\bigg\langle{#1}\bigg\rangle}
\newcommand{\x}{\mathbf{r}}
\newcommand{\bk}{\mathbf{k}}
\newcommand{\bp}{\mathbf{p}}
\newcommand{\re}{\mathrm{Re}}
\newcommand{\im}{\mathrm{Im}}

\newcommand{\ta}{t_{\rm asy}}
\newcommand{\tp}{t_{\rm pla}}

\author{Karol Gietka}
\affiliation{Faculty of Physics, University of Warsaw, ul. Pasteura 5, 02-093 Warsaw, Poland}
\author{Jan Chwede\'nczuk}
\affiliation{Faculty of Physics, University of Warsaw, ul. Pasteura 5, 02-093 Warsaw, Poland}
\author{Tomasz Wasak}
\affiliation{Max-Planck-Institut f\"ur Physik komplexer Systeme, 01187 Dresden, Germany}
\author{Francesco Piazza}
\affiliation{Max-Planck-Institut f\"ur Physik komplexer Systeme, 01187 Dresden, Germany}

\title{Multipartite-Entanglement Dynamics in Regular-to-Ergodic Transition:\\a Quantum-Fisher-Information approach}

\begin{abstract}
The characterization of entanglement is a central problem for the study of quantum many-body dynamics. Here, we propose the quantum Fisher information as a useful tool for the study of multipartite-entanglement dynamics in many-body systems.  We illustrate this by considering the regular-to-ergodic transition in the Dicke model---a fully-connected spin model showing quantum thermalization above a critical interaction strength. We show that the QFI has a rich dynamical behavior which drastically changes across the transition. In particular, the asymptotic value of the QFI, as well as its characteristic timescales, witness the transition both through their dependence on the interaction strength and through the scaling with the system size. Since the QFI also sets the ultimate bound for the precision of parameter estimation, it provides a metrological perspective on the characterization of entanglement dynamics in many-body systems. Here we show that quantum ergodic dynamics allows for a much faster production of metrologically useful states.
\end{abstract}

\maketitle

\section{Introduction}

A thorough understanding of quantum many-body dynamics necessarily requires the study of the time-evolution of the entanglement between the particles. In recent years, several studies have been performed especially in the context of thermalization in closed quantum many-body systems~\cite{PhysRevA.43.2046,PhysRevE.50.888,olshani,eisert_rev_2015,BORGONOVI20161,PhysRevLett.117.170404}. The computation of entanglement in strongly correlated systems is a challenging task, and a few general results are available for given system subclasses, like for the time-evolution of entanglement entropy in one-dimensional systems~\cite{calabrese2005evolution} and ergodic systems~\cite{kim_huse_2013,luitz_2016}, or the boundary-laws for asymptotic states of local Hamiltonians~\cite{osborne_2006}. The logarithmic growth in time of entanglement entropy has provided a clear distinction between Anderson- and many-body-localization~\cite{dechiara_2006,znidaric_2008,bardarson_mbl_2012,doi:10.1146/annurev-conmatphys-031214-014726,vasseur2016nonequilibrium}. While throughout these studies the entanglement is mostly characterized using bipartite entanglement entropies (which can be applied only to closed systems in pure quantum states), the interest in alternative measures has recently emerged~\cite{detomasi_mutual_2017,bera_2016,goold_2015,campbell2017dynamics}.

In this work, we propose the quantum Fisher information (QFI)~\cite{braunstein1994statistical,wootters1981statistical} as a useful quantity for the study of the time-evolution of entanglement in many-body systems. The QFI is a measure of genuine multipartite entanglement~\cite{pezze_entqfi_2009,hyllus_qfimult_2012}. With respect to entanglement entropies, it has the important advantage of being directly applicable to mixed states and thus to open-system dynamics. It is widely studied in the context of quantum metrology---as it sets a lower bound for the uncertainty in parameter estimation---but much less for the characterization of the dynamics of quantum many-body systems.  So far, the QFI has been used to detect phase transitions in ground or thermal states~\cite{hauke_qfisus_2016,ma_qfilmg_2009,wang2014quantum}, while its dynamical behavior across phase transitions remains largely unexplored, in particular in ergodic systems.

\begin{figure}[t]
  \centering
  \includegraphics[width=.45\textwidth]{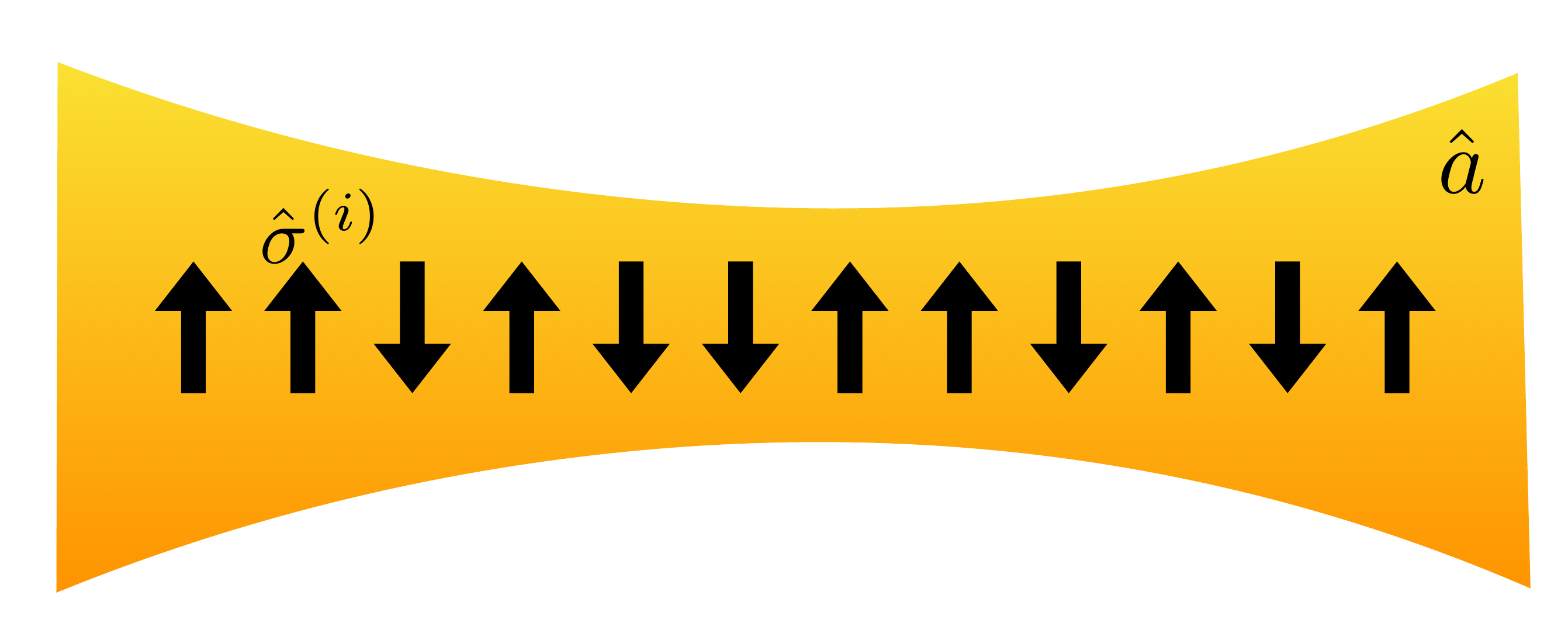}
  \caption{Graphical illustration of the Dicke Model (see Eq.~(\ref{eq:DM})) studied in this work. $N$ spins of length $1/2$ are all coupled to a single bosonic mode.}
\label{fig:DM_illustration}
\end{figure}

Here we show that the QFI provides a very rich characterization of quantum many-body dynamics across a regular to ergodic transition. We consider the Dicke model (DM), where interactions between $N$ spin-$1/2$ particles are mediated by a bosonic mode coupled at a rate $g$~\cite{dicke1954coherence,kirton2018introduction} (see also Fig.~\ref{fig:DM_illustration}). As shown by Altland and Haake~\cite{altland2012quantum,altland2012equilibration}, the Dicke model offers a paradigm for quantum thermalization dynamics with underlying classical chaos in a fully-connected system. In particular, the semiclassical dynamics in phase space shows a transition between regular and ergodic at a critical coupling strength $g_c$, consistently with the behavior of Hamiltonian spectrum turning from Poissonian to Wigner-Dyson level statistics~\cite{emary2003chaos}.

We find that the QFI in the DM has a dynamical behavior which drastically changes across the transition between the regular and the ergodic phase. Its asymptotic value, as well as the characteristic timescales, witness the transition both through their dependence on the control parameter $g$ and through their scaling with system's size~$N$.

\section{Summary of the main results}

Starting from a state which is not an eigenstate of the DM-Hamiltonian [see Eq.~(\ref{eq:DM})] and fixing the initial energy with respect to the ground state (see Fig.~\ref{fig:dos}), we compute the time evolution of the QFI optimized over all possible atomic single-particle operations.  In the regular phase $g<g_c$ (see Fig.~\ref{fig:QFI_summary} bottom row), the QFI shows oscillations around an envelope which grows continuously in time as $\mathrm{erf}( t^2/\ta^2)$ and asymptotically at $t\simeq\ta$ saturates to a value scaling at the Heisenberg limit (HL) $\propto N^2$. The latter is the strongest possible scaling for a system of $N$-qubits (see Fig.~\ref{fig:QFI_asympt_N}). On the other hand, in the ergodic phase $g>g_c$ (see Fig.~\ref{fig:QFI_summary} upper row) the QFI envelope first reaches an intermediate plateau with shot-noise (SN) scaling $\propto N$ within a time $\tp<\ta$, and only after $\ta$ reaches its final asymptotic value, the latter again showing Heisenberg-scaling. This double-step growth in the ergodic phase disappears at large enough initial energies (see Fig.~\ref{fig:QFI_summaryHE}), where the QFI envelope grows like $\mathrm{erf}( t^2/\ta^2)$ for all values of $g$.

While the asymptotic value of the QFI scales at the HL both in the ergodic and the regular phase, its actual value shows a sharp behavior at the critical coupling strength $g_c$, after which it rapidly grows (see blue points in Fig.~\ref{fig:QFI_asympt_g_combined}). Its behavior for $g=g_c$, however, becomes less sharp as energy is increased (see red points in Fig.~\ref{fig:QFI_asympt_g_combined}).

Also the timescale $\ta$ witnesses the transition as it abruptly drops by increasing $g$ down until $g_c$, after which is saturates to a constant value (see Fig.~\ref{fig:tasy_g_combined}). Remarkably, the behavior of $\ta$ as a function of $g$ is almost independent of the initial energy (compare red and blue points in Fig.~\ref{fig:tasy_g_combined}). Therefore, differently from the asymptotic value, the saturation time $\ta$ remains a good witness of the regular-to-ergodic transition at all energies.

The system-size dependence of the timescales is also clearly distinct in the two phases: in the regular phase $\ta\propto \sqrt{N}$, while in the ergodic phase the scaling of $\ta$ is consistent with $\log(N)$ at least at high-enough initial energies (see Fig.~\ref{fig:timescales_N_all}).  The fact that $\ta\propto\log(N)$ suggest its interpretation as the Ehrenfest time, which is proportional to the volume of the accessible phase space, the latter being proportional to $N$ in the DM~\cite{altland2012equilibration}.

This interpretation is confirmed by an analysis of the Wigner distribution function, according to which $\tp$ is connected to the formation of weakly squeezed nonclassical states while $\ta$ corresponds to the distribution fully covering the available region of phase space and forming small-scale structures of angular size $1/N$ (see Fig.~\ref{fig:wigner_erg_LE}).  The size of the region covered by the Wigner distribution quickly grows by increasing the initial energy until the whole phase space is taken (see Fig.~\ref{fig:wigner_erg_HE}), consistently with the prediction of Altland and Haake for the Husimi function~\cite{altland2012quantum} and the underlying classical chaos. As discussed above, at such initial energies the double-step growth of the QFI is absent. Such a double plateau formation can thus be related to the mixed character of the underlying classical phase space.

Upon a QFI-based characterization of many-body dynamics, one obtains a quantification of the usefulness of the given many-body state for quantum metrology, since the lower bound to the uncertainty of parameter estimation is set by the inverse square root of the QFI~\cite{helstrom1969quantum}. For instance, recent studies have demonstrated the metrological usefulness of quantum states generated during chaotic dynamics in the kicked top~\cite{fiderer2018quantum}. Our results here show that, compared to regular dynamics, quantum ergodic dynamics allows for a much faster production of entangled many-body states providing Heisenberg scaling of the metrological precision. 

\section{Quantum Fisher information and multipartite entanglement}
\label{sec:QFI_intro}
 
In this section, we will summarize the properties of the QFI which are relevant for the following analysis and discussion.

For a quantum state given by a density matrix $\hat\varrho$, the QFI is defined in relation to a chosen hermitian operator~$\hat{O}$, called the generator of the transformation, as
\begin{equation}\label{eq:QFIdef}
  I^Q[\hat\varrho;\hat{O}]=2\sum_{l,l'}\frac{\left(\lambda_l-\lambda_{l'}\right)^2}{\lambda_l+\lambda_{l'}}\big|\langle l|\hat{O}|l'\rangle\big|^2\ ,
\end{equation}
with the spectral decomposition of the density matrix given by $\hat\varrho=\sum_l\lambda_l|l\rangle\langle l|$, where $\lambda_l>0$ and $\sum_l\lambda_l=1$. For pure states, the QFI reduces to $I^Q=4(\Delta\hat{O})^2=4(\mathrm{Tr}[\hat\varrho\hat{O}^2]-\mathrm{Tr}[\hat\varrho\hat{O}]^2)$, i.e., four times the variance of the operator.  The QFI can be interpreted as a square of a ``statistical speed''~\cite{braunstein1994statistical,wootters1981statistical}, defined as the rate of change of the absolute statistical distance between two quantum states along a single-parameter path generated by the operator $\hat{O}$ through

\begin{align}\label{eq:tr}
  \hat\varrho(\theta)=e^{-i\theta\hat{O}}\hat\varrho e^{i\theta\hat{O}}.
\end{align}
For thermal states, the QFI coincides with the dynamic susceptibility related to the operator $\hat{O}$~\cite{hauke_qfisus_2016}.

In our work, the QFI is used as a mean to characterize multipartite entanglement. In the particular case of the DM considered here [see Eq.~(\ref{eq:DM})], we have a composite Hilbert space made of a bosonic degree of freedom (L) and a spin sub-system (S) made of $N$ spins of length $1/2$.  The total Hilbert space is a tensor product of the two sub-spaces

\begin{align}
  \mathcal H=\mathcal H_{\mathrm L}\otimes\mathcal H_{\mathrm S}.
\end{align}
We focus on pure states of the composite system, i.e., $\ket\Psi\in\mathcal H$ and concentrate on the multipartite entanglement in the spin sub-space. We consider the following {\it linear} (in a sense that no products of $\hat{\vec{\sigma}}^{(l)}$ and $\hat{\vec{\sigma}}^{(l')}$ appear) operators
\begin{align}
\label{eq:linop_general}
\hat{O}_{\rm lin}=\hat{\mathbb I}_{\mathrm L}\otimes \frac12\sum_{l=1}^N\vec{n}^{(l)}\cdot\hat{\vec{\sigma}}^{(l)},
\end{align}
with the Pauli operators $\hat{\vec{\sigma}}^{(l)}=(\hat\sigma_x^{(l)},\hat\sigma_y^{(l)},\hat\sigma_z^{(l)})$ and the vectors $\vec{n}^{(l)}=(n_x ^{(l)},n_y ^{(l)},n_z ^{(l)})$ that define the rotation of the Bloch sphere, such that $(n_x ^{(l)})^2+(n_y ^{(l)})^2+(n_z ^{(l)})^2=1$.  The only possible pure state $\ket\Psi$ in which the spins and bosons are non-entangled is of the form
\begin{align}
  \ket\Psi=\ket\psi_{\mathrm L}\otimes\ket{\phi_1}\otimes\ket{\phi_2}\cdots\otimes\ket{\phi_N},
\end{align}
where $\ket\psi_{\mathrm L}$ is a state of the bosonic subsystem and $\ket{\phi_i}$ is a state of the $i$th single spin. If all the spins are in the same state, the $N$-body state $\ket\phi^{\otimes N}$ is called the coherent spin state (CSS). For non-entangled states and the transformation~(\ref{eq:linop_general}), the QFI is bounded by
\begin{align}
  I^Q[\ket\Psi;\hat{O}_{\rm lin}]=4(\Delta\hat{O}_\mathrm{lin})^2\leqslant N,
\end{align}
i.e., the shot-noise limit (SNL)~\cite{pezze_entqfi_2009}. This bound can be overcome when the spins are entangled. To see this, consider an exemplary state
\begin{align}\label{eq:ex}
  \ket\Psi=\frac1{\sqrt2}\left(\ket0\otimes\ket{\uparrow}^{\otimes N}+\ket1\otimes\ket{\downarrow}^{\otimes N}\right),
\end{align}
where $\ket 0$ and $\ket 1$ are a bosonic vacuum and a one-particle state, respectively, and $\ket{\uparrow}/\ket{\downarrow}$ are the eigenstates of $\hat\sigma_z$ with eigenvalues $\pm1$. Taking now
\begin{align}
\label{eq:linopDM}
  \hat{O}_{\rm lin}=\hat{\mathbb I}_{\mathrm L}\otimes \hat J_z,
\end{align}
with the collective angular momentum operators of $N$ spins defined as
\begin{align}
\label{eq:coll_angmom}
  \hat J_{\vec{n}}=\frac12\vec{n}\cdot\left(\sum_{l=1}^N\hat{\vec{\sigma}}^{(l)}\right)
\end{align}
we obtain
\begin{align}
  I^Q[\ket\Psi;\hat{O}_{\rm lin}]=N^2,
\end{align}
which is the Heisenberg limit (HL) and is the maximal value of the QFI for the family of linear transformations (\ref{eq:linop_general}).  The example (\ref{eq:ex}) can also be used to illustrate one additional property of the QFI for transformations (\ref{eq:linop_general}).  Taking the reduced density matrix of spins, obtained after tracing out the bosonic degree of freedom from the state~(\ref{eq:ex}), we obtain
\begin{align}\label{eq:ex_trace}
  \hat\varrho_{\rm S}=\tr{\ketbra\Psi\Psi}_{\rm L}=\frac12\Big((\ketbra\uparrow\uparrow)^{\otimes N}+(\ketbra\downarrow\downarrow)^{\otimes N}\Big).
\end{align}
This is a separable (non-entangled) state, for which the QFI calculated from Eq.~(\ref{eq:QFIdef}) gives $I^Q\leqslant N$ for any $\hat J_{\vec n}$, thus the SNL~\cite{pezze_entqfi_2009}.  

To summarize, the QFI calculated with the operators~(\ref{eq:linop_general}) is more than just a criterion for the entanglement between spins. It rather detects the entanglement within the spin sub-space of the full many-body state $\ket\Psi\in\mathcal H$, that is either we have entangled spins classically correlated with the bosons, or non-entangled spins non-classically correlated with bosons.

We stress that the QFI is an experimentally accessible quantity even in systems of many qubits, as recently demonstrated with ultracold atoms~\cite{Luecke773,Strobel424}. Also in view of the recent proposals for an efficient QFI witnessing~\cite{PhysRevLett.116.090801,Pezze11459,PhysRevA.95.032330}, the extensions to more complex and larger systems seems a concrete possibility in the near future.

For the following analysis of multipartite-entanglement dynamics, we will compute the QFI for the momentary quantum state and maximize it at every instant in time with respect to all possible operators of the class (\ref{eq:linop_general}). 

As the DM-Hamiltonian (\ref{eq:DM}) is fully connected and thus contains only collective angular-momentum operators of the type~(\ref{eq:coll_angmom}), it is sufficient to consider the following optimized QFI
\begin{align}\label{eq:opt}
  I^Q(t)=\underset{\vec{n}}{\mathrm{max}}\ I^Q[\hat\varrho(t);\hat{\mathbb{I}}_{\rm L}\otimes J_{\vec{n}}  ]\ ,
\end{align}
which reduces the maximization problem to finding the optimal vector $\vec{n}$ defining the rotation axis.  Here $\hat\varrho(t)=\hat U(t)\hat\varrho_0\hat U^\dag(t)$, $\hat\varrho_0$ being the initial state and $\hat U(t)$ the unitary evolution operator generated by the Hamiltonian.

We note that $I^Q(t)$ is {\it not} related to an echo fidelity~\cite{gorin2006dynamics}---an important quantity in the context of thermalization and irreversibility which has been studied also for the Dicke model~\cite{prosen2003evolution}---since in our case the path in density-matrix space is not generated by the time-evolution operator $\hat U(t)$ but rather by $\hat{O}_{\rm lin}$ at every instant in time.

Finally, we point out that the QFI also provides ultimate bounds for the precision of estimation of a metrological parameter $\theta$ under the transformation (\ref{eq:tr}).  Accordingly, the bound for the uncertainty of the parameter estimation is $\Delta\theta\geqslant I^Q[\hat\varrho;\hat{O}_\mathrm{lin}]^{-1/2}$. Separable states at most achieve the SNL sensitivity while maximally entangled states can in principle yield the HL precision~\cite{helstrom1969quantum,braunstein1994statistical,pezze_entqfi_2009}.

\section{Model and approach}
\label{sec:DM_intro}

\begin{figure}[t]
  \centering
  \includegraphics[width=\columnwidth]{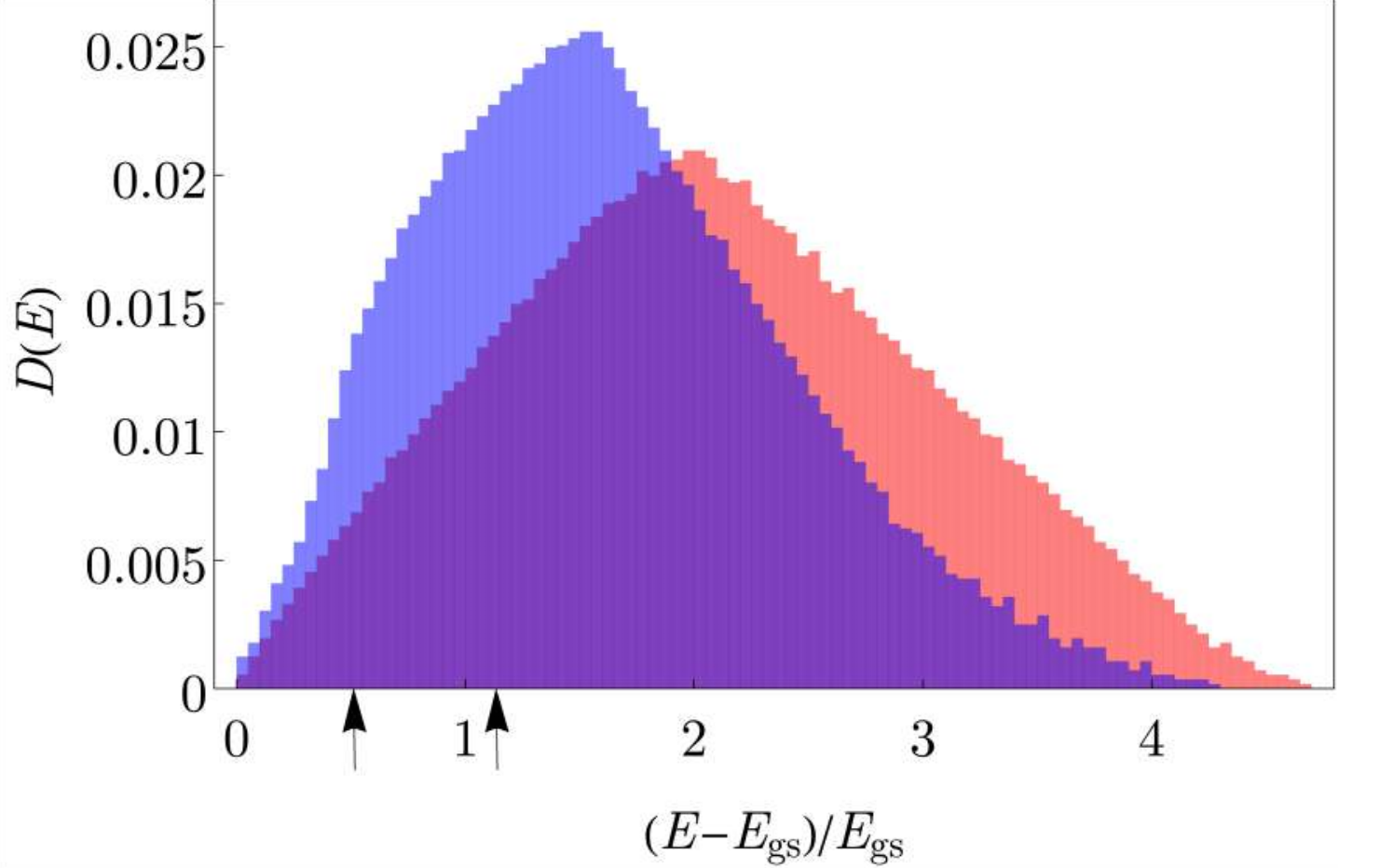}
  \caption{Density of states as a function of $(E-E_{\mathrm{gs}})/E_{\mathrm{gs}}$ for the Hamiltonian (\ref{eq:DM}) for $N=100$ and $g = 0.3$ (red) or $g = 0.9$ (blue). Arrows indicate the two different initial energies which are compared throughout the following analysis. The density of states on the right of the peak converges slowly with the cutoff in the bosonic Fock space. In our case, we choose it such that the density of states converged in the whole region on the left of the peak. This guarantees that our numerical results are independent of the cutoff.}
\label{fig:dos}
\end{figure}

\begin{figure*}[htp]
  \centering
  \includegraphics[width=.9\textwidth]{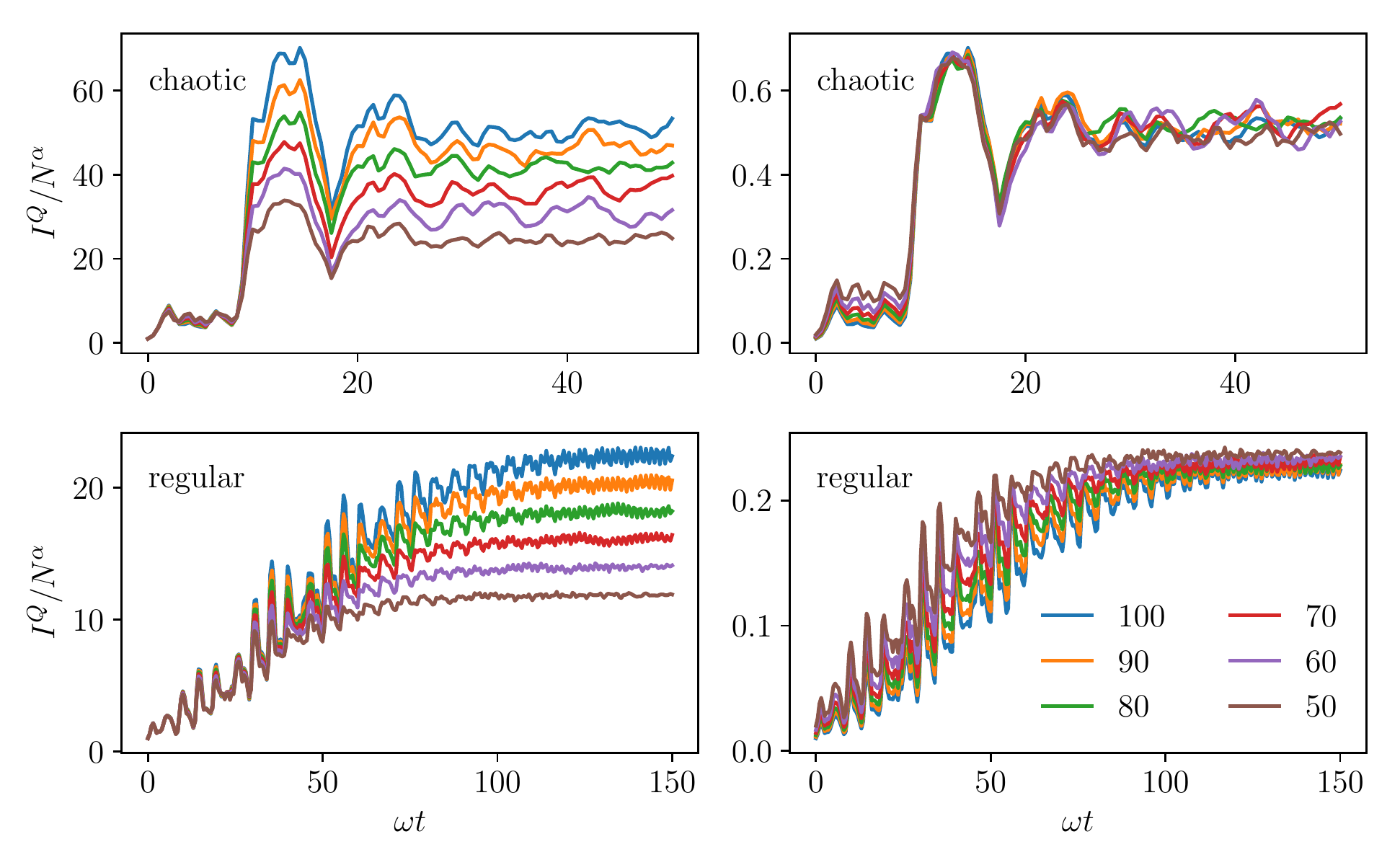}
  \caption{Dynamics of the QFI in the chaotic (top row) vs. regular phase (bottom row). To render the different scaling regimes apparent, the QFI $I^{Q}/N^{\alpha}$ is shown for different system sizes $N$ (ranging from $50$ to $100$ with $\Delta N = 10$) for $\alpha=1$ (left column) and $\alpha=2$ (right column). Hereinafter, the parameters are expressed in units of $\omega$, and the energy of the initial state is fixed with respect to the ground state, i.e., $(E-E_{\mathrm{gs}})/E_{\mathrm{gs}} = 0.53$ (see also Fig.~\ref{fig:dos}). Here $\omega_0 = 1$, $g = 0.9$ (top row) and $g = 0.4$ (bottom row).}
\label{fig:QFI_summary}
\end{figure*}

A paradigmatic feature of generic many-body systems is ergodicity, that is, the ability to relax to an asymptotic state which can be effectively described by thermal-equilibrium. For closed quantum systems the issue of thermalization---and its absence for instance due to integrability or disorder---is still an object of intensive study~\cite{PhysRevA.43.2046,PhysRevE.50.888,olshani,eisert_rev_2015,BORGONOVI20161,PhysRevLett.117.170404}. With the aim of putting forward the QFI for the characterization of many-body dynamics, we consider here the Dicke model~\cite{dicke1954coherence,kirton2018introduction}, which is one of the simplest instances of a quantum many-body system showing thermalization. Its Hamiltonian is given by (setting $\hbar=1$): 
\begin{align} 
    \hat H = \omega_0 \hat J_z + \omega \, \hat
  a^\dagger \hat a +\frac{2g}{\sqrt{N}}\left(\hat a +\hat
  a^\dagger\right)\hat J_x \label{eq:DM}
\end{align}
The simplicity of the DM resides in it being a fully-connected model---only the collective spin operators $\hat J_{\vec{n}}$ appear in Eq.~\eqref{eq:DM}---with interactions between the $N$ spin-$1/2$ particles mediated by a single bosonic degree of freedom $\hat{a}$ satisfying $\left[ \hat a, \hat a^\dagger \right]=1$. 

The DM features a second-order phase transition in the ground state at the critical coupling $g_c=0.5$, above which the average value of the total magnetization $\av{\hat J_x}$ as well as the coherent contribution to the bosonic field $\av{\hat a}$ become finite, thereby spontaneously breaking the $Z_{\mathbf{2}}$-symmetry of the Hamiltonian \eqref{eq:DM}: $\hat{a}\to -\hat{a}, \hat{J}_x\to -\hat{J}_x$.  What is more relevant for the present work is that at $g=g_c$ also the whole spectrum of eigenstates changes: the level statistics namely shows a transition from Poissonian below $g_c$ to Wigner-Dyson above~\cite{emary2003chaos}. This indicates that the DM should behave ergodically above $g_c$. Indeed, this is confirmed by the semiclassical analysis of~\cite{altland2012quantum,altland2012equilibration}, showing that in this case a state initially localized in phase space eventually covers homogeneously the whole phase space available at the initially fixed energy, that is, the system relaxes to a microcanonical distribution. A semiclassical study of the DM is justified as a perturbative expansion in $1/N$ since the fully connected nature makes it such that the DM possesses a classical limit for $N\to\infty$. The classical dynamics shows a crossover from regular to chaotic which can be connected to the thermalizing behavior of the full quantum model.

Our aim is to study the entanglement dynamics in the DM at finite $N$ using the QFI, with particular attention to the characterization of the transition from regular to ergodic behavior as a function of $g$. 

\section{Results}

\begin{figure*}[htp]
  \centering
  \includegraphics[width=.9\textwidth]{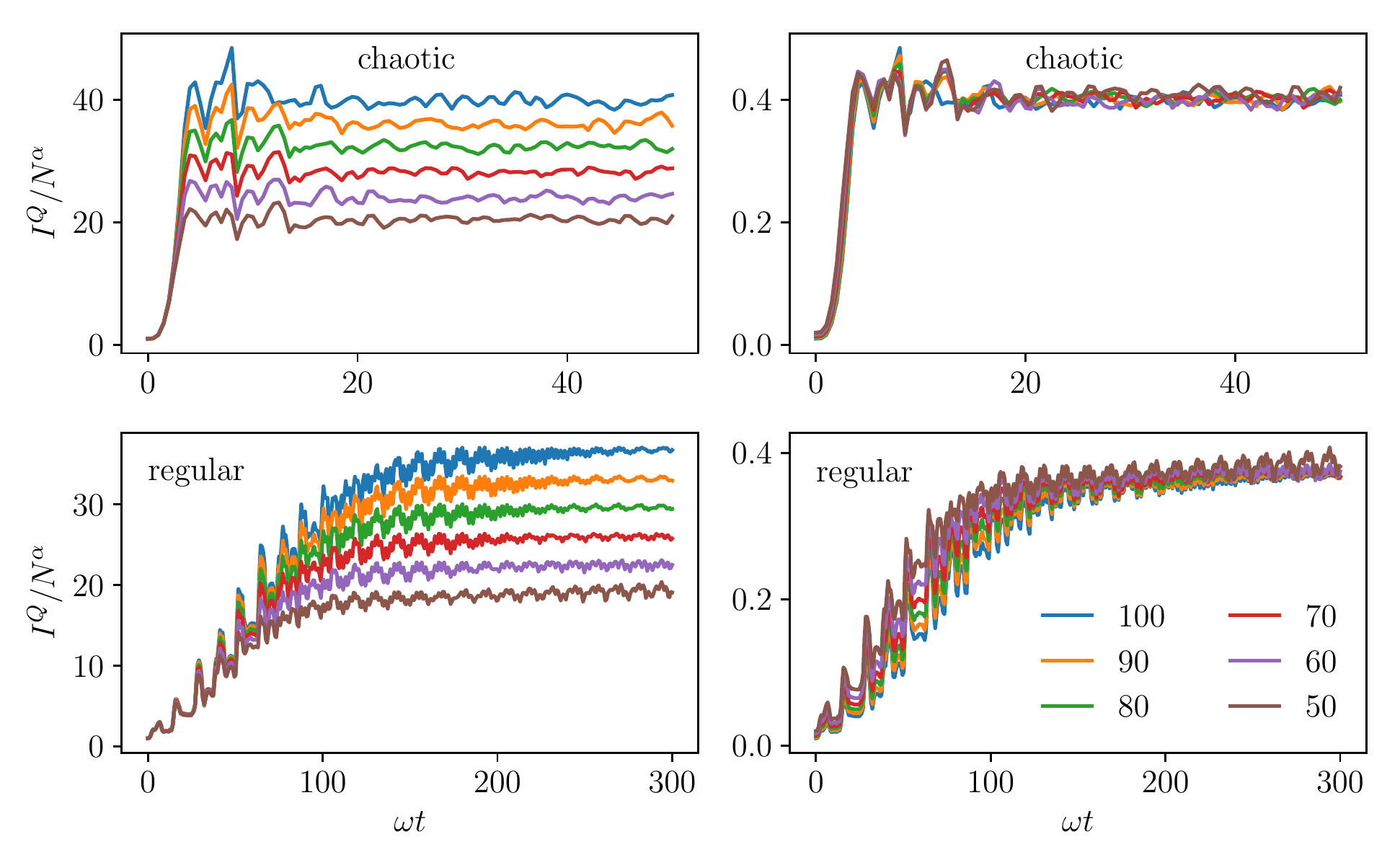}
  \caption{Same as in Fig.~\ref{fig:QFI_summary}, this time for a higher initial energy $(E-E_{\mathrm{gs}})/E_{\mathrm{gs}}= 1.11$ (see also Fig.~\ref{fig:dos}).}
\label{fig:QFI_summaryHE}
\end{figure*}

We compute the time evolution of the optimized QFI $I^Q(t)$ defined in Eq.~(\ref{eq:opt}) by starting from an initial pure state $\ket\Psi_0$ with an initial energy $E=\bra{\Psi_0}\hat{H}\ket{\Psi_0}$. While changing the coupling strength $g$ and the number of spins $N$, we keep the ratio of the initial energy to the ground state energy $E_{\rm gs}$ fixed. In the following, we present results for two different energy ratios corresponding to the arrows shown in Fig.~\ref{fig:dos}. We pick the initial state to be $\ket\Psi_0=\ket\alpha\otimes\ket{\phi}^{\otimes N}$ i.e. the product of a coherent state of the bosons $\hat{a}\ket\alpha=\alpha\ket\alpha$ and a CSS for the spins. This is not an eigenstate of the DM-Hamiltonian (\ref{eq:DM}), and we fix its average energy by choosing  the CSS of spins and adjusting the value of $\alpha$.

\subsection{QFI dynamics}

The time evolution of the optimized QFI is shown in Figs.~\ref{fig:QFI_summary} and~\ref{fig:QFI_summaryHE} for the two different initial energies. In each figure, we compare the typical dynamics below and above $g_c$. Based on the behavior of the level statistics~\cite{emary2003chaos} discussed in section~\ref{sec:DM_intro}, we will refer to the parameter region $g<g_c$ as the regular phase and to the region $g>g_c$ as the ergodic phase. In all cases, the QFI shows oscillations around an envelope, the latter growing in time until it reaches a stationary asymptotic value $I_\infty^Q$ within a timescale $\ta$. In the regular phase (bottom row of Figs.~\ref{fig:QFI_summary} and~\ref{fig:QFI_summaryHE}), the QFI envelope grows steadily and is well fitted by
\begin{align}
\label{eq:QFI_fit}
I_{\rm env}^Q(t)=I_0^Q+(I_\infty^Q-I_0^Q)\,\mathrm{erf}\bigg(\frac{t^2}{\ta^2}\bigg),
\end{align}
where $\mathrm{erf}$ is the error function. On the other hand, in the ergodic phase the QFI shows a two-step growth, first reaching an intermediate plateau within a time $\tp$ and then suddenly abandoning it to reach its asymptotic value for $t>\ta$. This two-step growth, however, disappears at high enough energies, as shown in Fig.~\ref{fig:QFI_summaryHE}, where the intermediate plateau is absent, and the QFI envelope is always well fitted by the functional form (\ref{eq:QFI_fit}).

\subsection{Asymptotic value}

\begin{figure}[b]
  \centering
  \includegraphics[width=.5\textwidth]{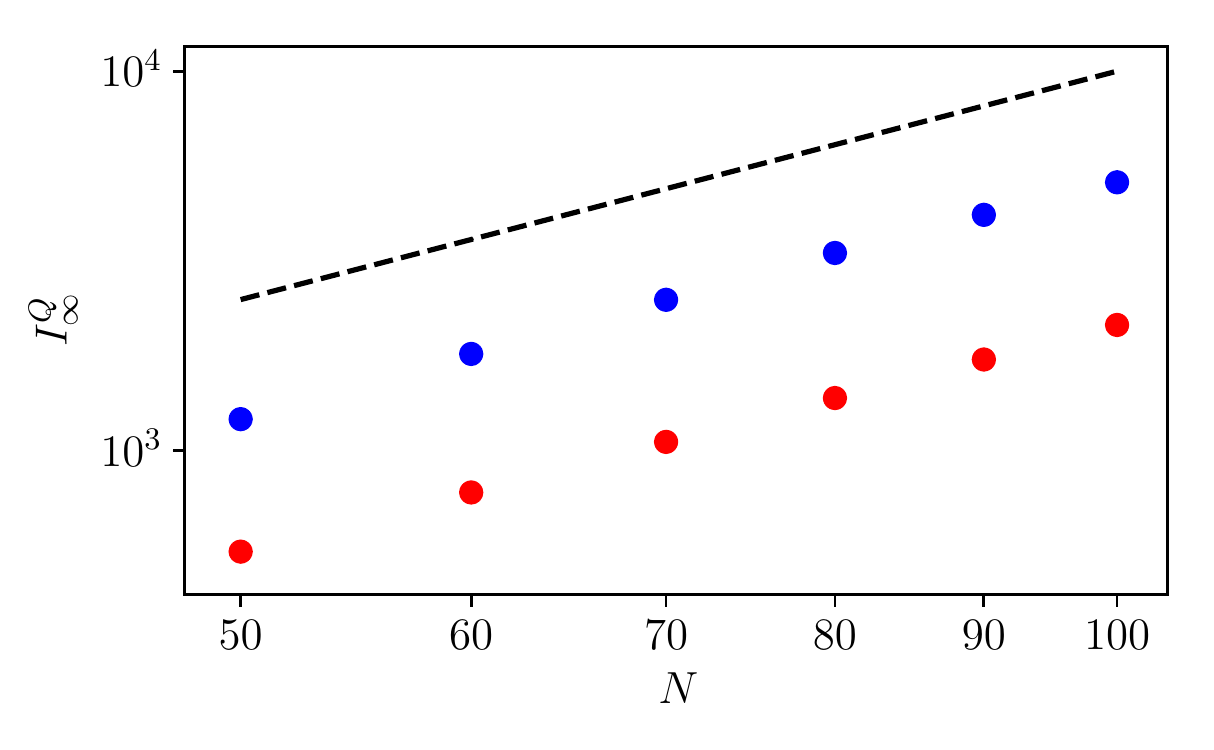}
  \caption{Asymptotic value $I_{\infty}^Q$ of the QFI as a function of $N$ in double-logarithmic scale for $g = 0.9$ (blue curve) and $g=0.4$ (red curve). The black dashed curve is the HL, i.e., $I_\infty^Q = N^2$. Here, the initial energy is $(E-E_{\mathrm{gs}})/E_{\mathrm{gs}} = 0.53$.}
\label{fig:QFI_asympt_N}
\end{figure}

In Figs.~\ref{fig:QFI_summary} and~\ref{fig:QFI_summaryHE}, we present two different rescalings of the QFI: the left panels show $I^Q/N$ while the right panels show $I^Q/N^2$. First, we observe that the asymptotic value $I_\infty^Q$ always scales like $N^2$ as indicated by the overlapping curves for $t>\ta$ in all the right panels, independently of the coupling strength and the initial energy. This is shown more directly in Fig.~\ref{fig:QFI_asympt_N}, where $I_\infty^Q$ is plotted as a function of $N$ both in the regular and the ergodic phase. We note that $I_\infty^Q$ scales with $N$ like the HL but lies below the ultimate bound by a numerical prefactor $\sim 1/2$. On the other hand, the upper-left panel of Fig.~\ref{fig:QFI_summary} shows that the QFI in the intermediate plateau appearing for $\tp<t<\ta$ scales like $N$, i.e., like the SNL.

\begin{figure}[t]
  \centering
    \includegraphics[width=.5\textwidth]{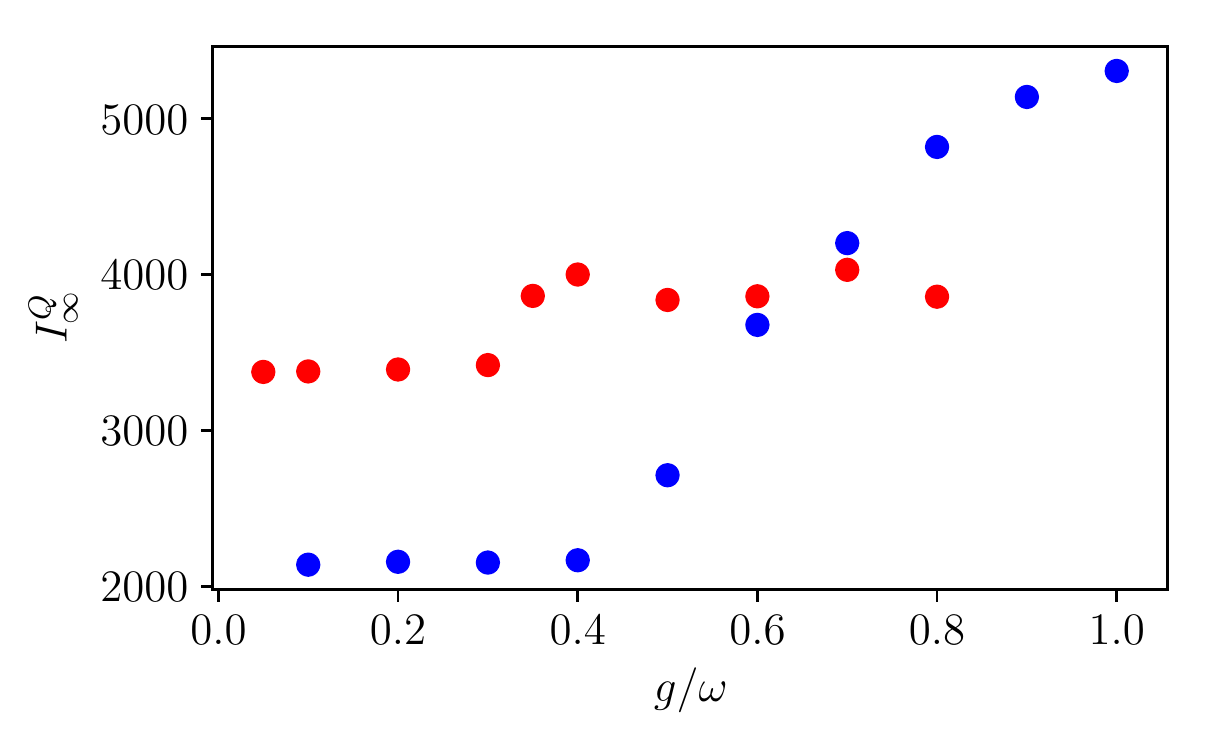}
  \caption{Asymptotic value $I_{\infty}^Q$ of the QFI across the regular-to-ergodic transition. Here $N=100$, and the initial energy $(E-E_{\mathrm{gs}})/E_{\mathrm{gs}} = 0.53$ (blue) or $(E-E_{\mathrm{gs}})/E_{\mathrm{gs}}= 1.11$
    (red).}
\label{fig:QFI_asympt_g_combined}
\end{figure}

The scaling of the asymptotic value $I_\infty^Q$ with the system's size $N$ does not allow to distinguish the regular from the ergodic phase. However, the behavior of $I_\infty^Q$ as a function of the coupling strength $g$ can much better distinguish the two phases. As shown in Fig.~\ref{fig:QFI_asympt_g_combined}, for $(E-E_{\mathrm{gs}})/E_{\mathrm{gs}} = 0.53$ the asymptotic value shows a sharp transition at $g=g_c=0.5\omega$. $I_\infty^Q$ is namely almost constant below $g_c$ and suddenly grows above. This behavior however becomes less and less sharp as the initial energy grows, as testified by the red points in Fig.~\ref{fig:QFI_asympt_g_combined} at $(E-E_{\mathrm{gs}})/E_{\mathrm{gs}} = 1.11$. By increasing the initial energy, not only the value of $I_\infty^Q$ increases in the regular phase, but it also decreases in the ergodic phase. Moreover, the value of $g$ at which $I_\infty^Q$ starts appreciably growing is slightly moved to lower $g$~\footnote{The fact that the asymptotic value of the QFI is larger in the ergodic phase was already observed in [\onlinecite{song2012}].}. 

\subsection{Characteristic timescales}

\begin{figure}[t]
  \centering
  \includegraphics[width=.5\textwidth]{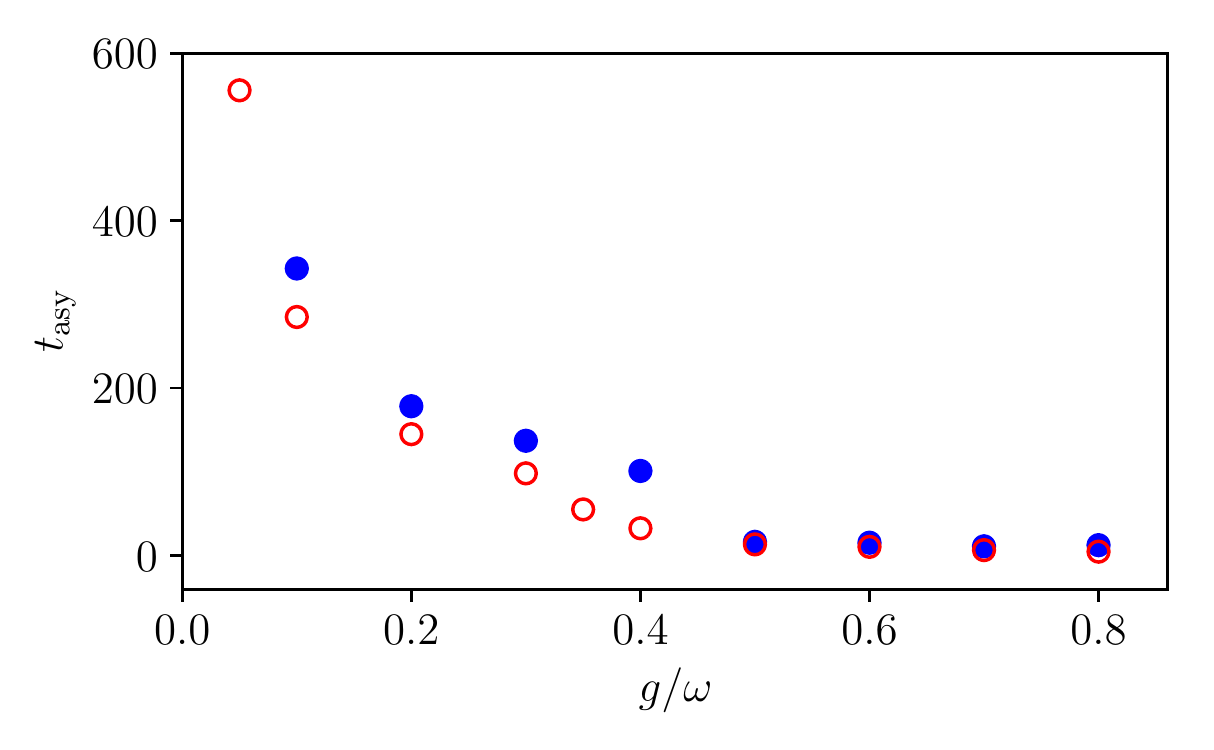}
  \caption{Behavior of the saturation time $t_{asy}$ across the regular-to-ergodic transition. Here $N=100$, and the initial energy $(E-E_{\mathrm{gs}})/E_{\mathrm{gs}} = 0.53$ (blue) or $(E-E_{\mathrm{gs}})/E_{\mathrm{gs}}= 1.11$ (red).}
\label{fig:tasy_g_combined}
\end{figure}

The timescales characterizing the dynamics of the QFI constitute an even better witness of the regular-to-ergodic transition. As shown in Fig.~\ref{fig:tasy_g_combined}, the time $\ta$ required for the QFI to reach its asymptotic value quickly decreases by increasing the coupling strength $g$ until the latter reaches $g_c$~\footnote{In the case of the double plateau, the fit starts from the fixed point at the end of the first plateau.}. Upon entering the ergodic phase for $g>g_c$, $\ta$ settles to an essentially constant value. Remarkably, this sharp behavior across $g_c$ is present independent of the initial energy, as apparent from comparing the red and blue circles in Fig.~\ref{fig:tasy_g_combined}.  In particular, the fact that the time required for the multipartite entanglement measured by the QFI to saturate does not depend on the interaction strength seems a good indicator for the ergodic character of the system.

\begin{figure}[b]
  \centering
  \includegraphics[width=.5\textwidth]{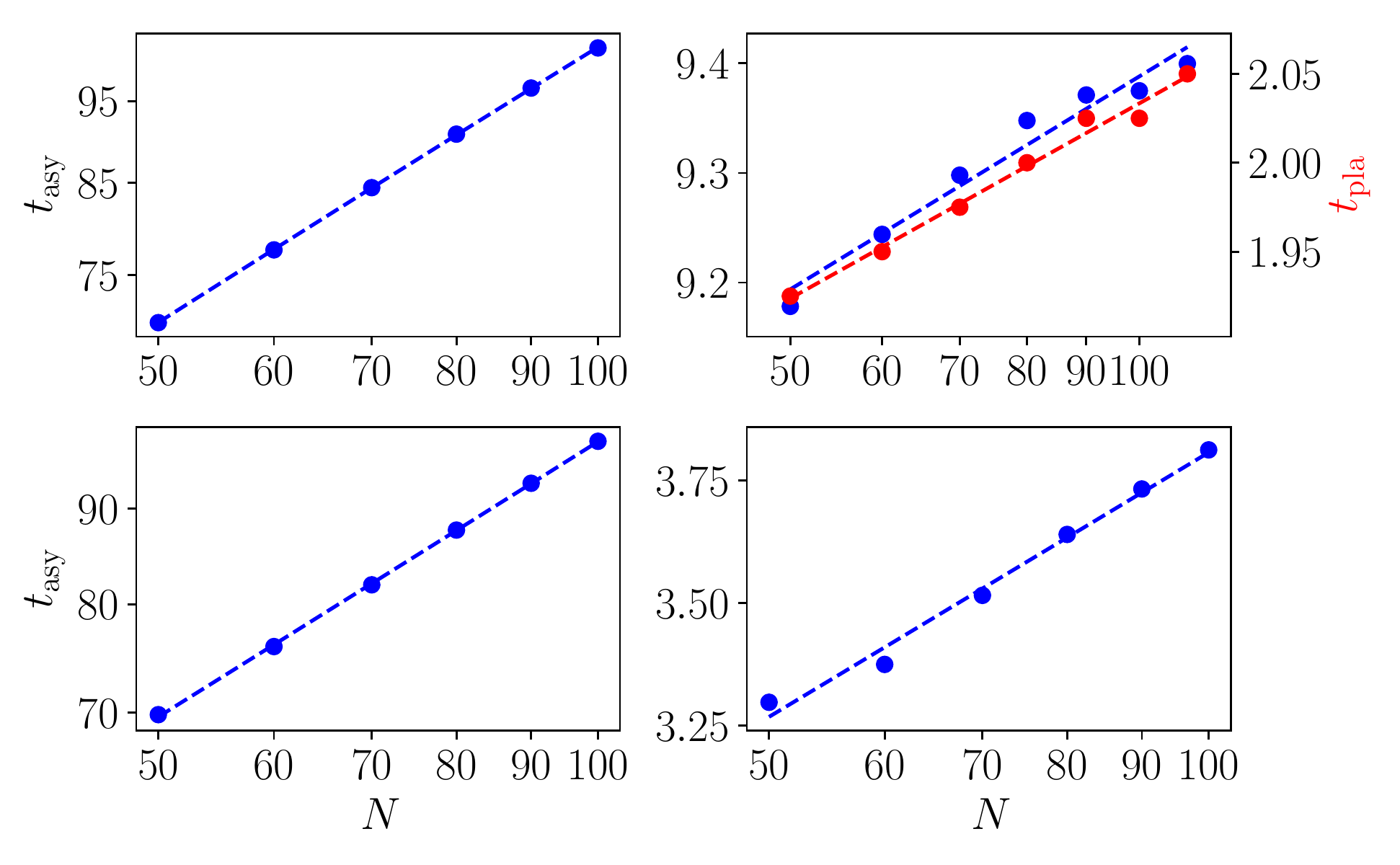}
  \caption{Behavior of the timescales as a function of system size in the regular ($g=0.3$, left column, double-logarithmic scale) vs. ergodic phase (here $g=0.9$, right column, semi-log scale) for initial energies  $(E-E_{\mathrm{gs}})/E_{\mathrm{gs}} = 0.53$ (upper row) and $(E-E_{\mathrm{gs}})/E_{\mathrm{gs}}= 1.11$ (bottom row). Dashed lines are guides to the eye. In the regular phase the asymptotic time $\ta$ is well fitted by $\sqrt{N}$. In the the ergodic phase the scaling is consistent with $\log(N)$ for $(E-E_{\mathrm{gs}})/E_{\mathrm{gs}}= 1.11$. This is less clear at $(E-E_{\mathrm{gs}})/E_{\mathrm{gs}} = 0.53$, where we also show the scaling of the intermediate-plateau time $\tp$.}
\label{fig:timescales_N_all}
\end{figure}

The regular and the ergodic phase can also be distinguished by the scaling of the saturation time $\ta$ with the system's size $N$, as shown in Fig.~\ref{fig:timescales_N_all}. In the regular phase, the dependence of $\ta$ on $N$ is very well fitted by $\sqrt{N}$. This holds independently of the initial energy, as one can see in the left panels of Fig.~\ref{fig:timescales_N_all}. In the ergodic phase instead the scaling of $\ta$ is consistent with $\log(N)$ (see lower right panel of Fig.~\ref{fig:timescales_N_all}).  On a qualitative level, this implies that in approaching the thermodynamic limit with our fully-connected model, the time required to reach the asymptotic, HL-scaling value of the entanglement diverges much slower with system size in the ergodic phase.

On the other hand, in the ergodic phase but at low enough energies we have seen that an intermediate plateau appears between $\tp$ and $\ta$. In the upper right panel of Fig.~\ref{fig:timescales_N_all} we see that in this case the scaling of $\ta$ and $\tp$ with the system's size is not as well fitted by $\log(N)$, at least for the sizes we explore here. This might be due to the mixed nature of the underlying
  classical phase space, as we discuss in the next section.
  
\subsection{Wigner distribution}

\begin{figure}[t]
  \centering
  \includegraphics[width=.4\textwidth]{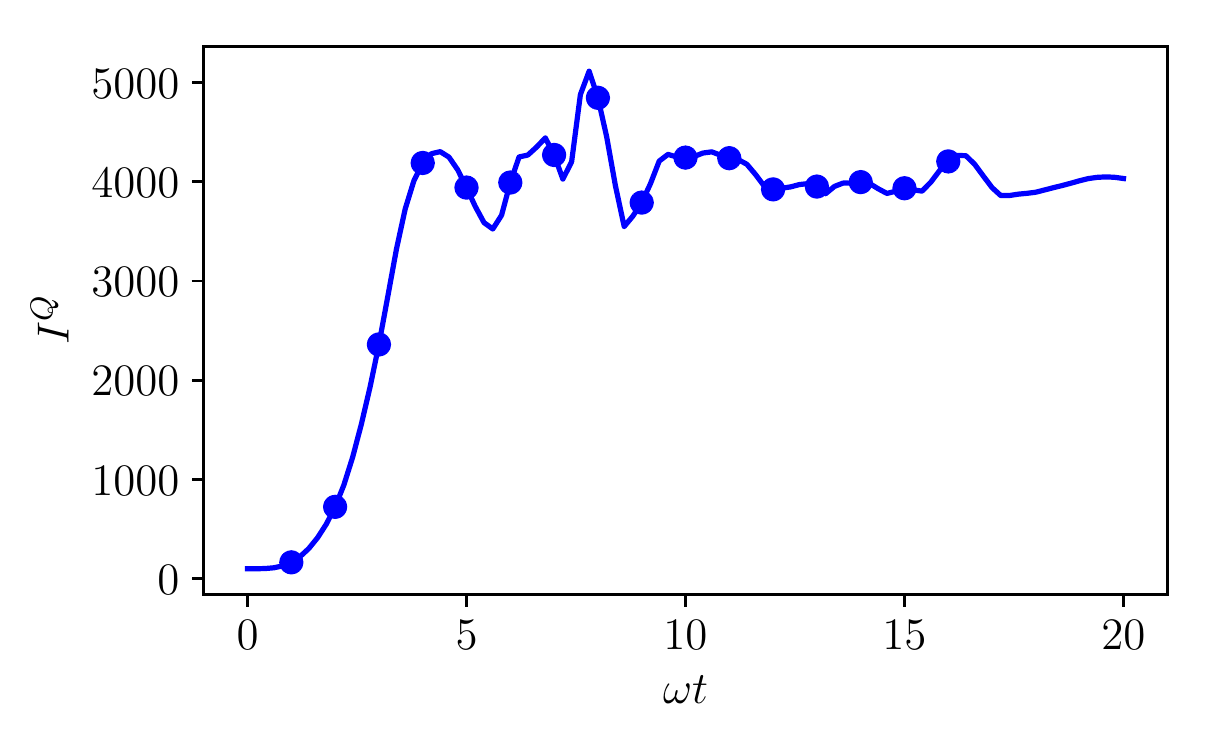}
  \includegraphics[width=.5\textwidth]{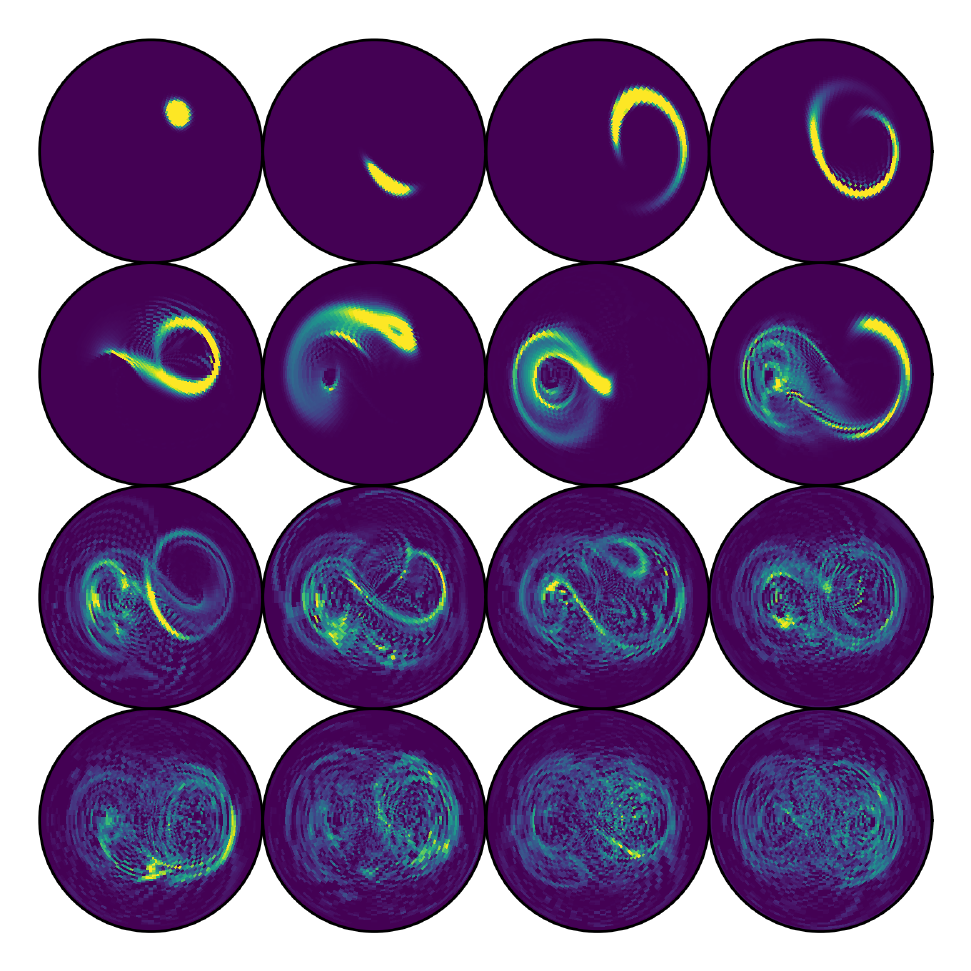}
  \caption{Time-evolution of the Wigner phase-space distribution in the ergodic phase at large energies. Points on the $I^Q$ curve are the points at which we calculate the Wigner function of the state. Here $N=100$ and the remaining parameters are the same as in Fig.~\ref{fig:QFI_summaryHE}, i.e., $(E - E_\mathrm{gs})/E_\mathrm{gs} = 1.11$, $g=0.9$.}
\label{fig:wigner_erg_HE}
\end{figure}

\begin{figure}[t]
  \centering
  \includegraphics[width=.4\textwidth]{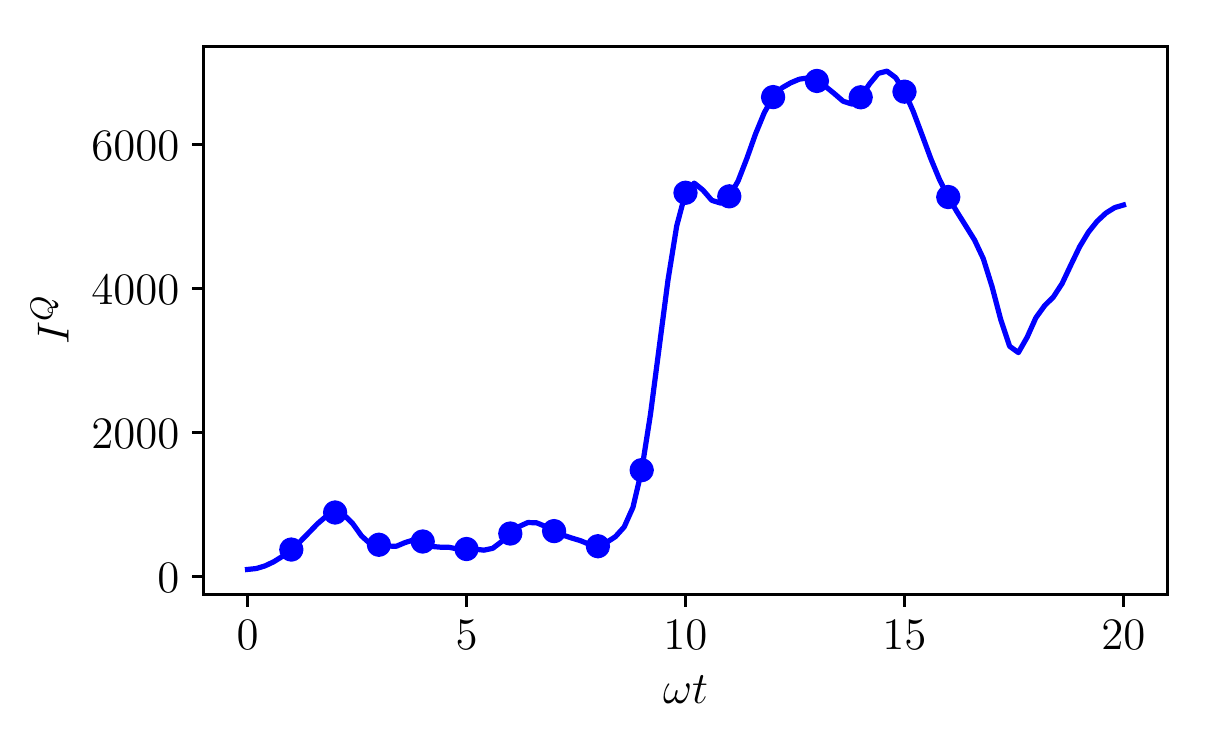}
  \includegraphics[width=.5\textwidth]{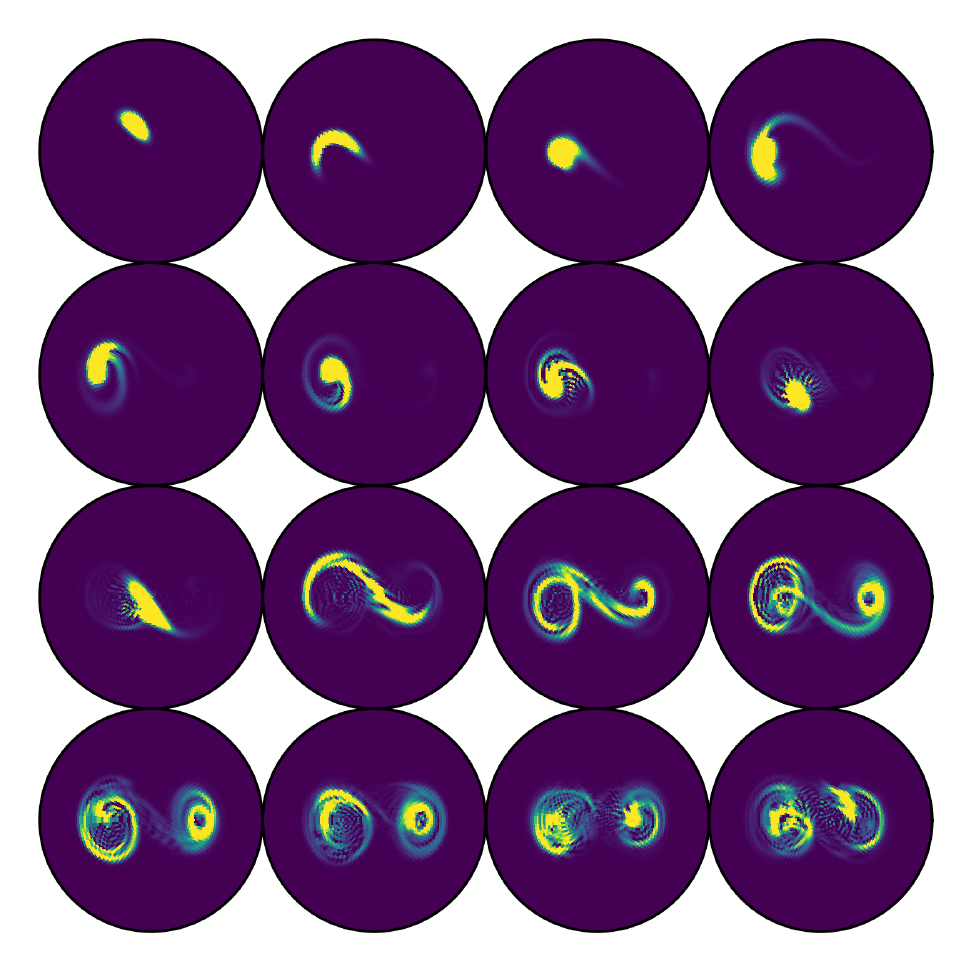}
  \caption{The same as in Fig.~\ref{fig:wigner_erg_HE}, except that now there parameters are: $(E - E_\mathrm{gs})/E_\mathrm{gs} = 0.53$ and $g=0.9$  (cf. Fig.~\ref{fig:QFI_summary}).}
\label{fig:wigner_erg_LE}
\end{figure}

The log-scaling with the system size of the saturation time $\ta$ in the ergodic phase suggests an interpretation as an Ehrenfest time. The latter is related to the breakdown of the semiclassical description of the dynamics, and the time at which this happens is known to scale as the logarithm of the volume of the available phase-space. This in turn for the DM depends linearly on $N$, and so the Ehrenfest time scales as $\log N$~\cite{altland2012equilibration}. 

We validate this hypothesis by analyzing the dynamics of the $SU(2)$ Wigner distribution, defined by~\cite{Agarwal1981}:
\begin{align}
    W(\theta,\phi) =\sum_{k=0}^{2j}\sum_{q=-k}^k Y_{kq}(\theta,\phi)G_{kq},
\end{align}
where $Y_{kq}$ are the spherical harmonics, and $G_{kq}$ are expansion coefficients in the basis of multipole operators $\hat T_{kq}$~\cite{edmonds1955angular} of the reduced density matrix for the spin subsystem $\hat \varrho_{\mathrm{S}}(t) \equiv \mathrm{Tr\left[\Psi \rangle \langle \Psi|\right]_{\mathrm{L}}}$:
\begin{align}
    \hat \varrho_{\mathrm{S}} = \sum_{k=0}^{2j}\sum_{q=-k}^k  G_{kq}\hat T_{kq}.
\end{align}
The above Wigner function is defined on the phase space of the spin degrees of freedom spanned by two angles $\theta,\phi$. 

The asymptotic plateau reached after $\ta$ is characterized by the QFI scaling like the HL $\propto N^2$, i.e., maximal entanglement (reduced by a prefactor $\sim1/2$, see Fig.~\ref{fig:QFI_asympt_N}). Correspondingly, the Wigner function shown in Fig.~\ref{fig:wigner_erg_HE} (in azimuthal equidistant projection) quickly spreads over a larger portion of the phase space, ultimately covering it fully for high-enough initial energy.  While spreading over phase space, the Wigner function forms small-scale structures of characteristic size $1/N$, as expected from ergodic quantum systems~\cite{zurek_2001}. This small-scale structures in phase space are responsible for the scaling with $N^2$ of the QFI~\cite{pezze_entqfi_2009}.

On the other hand, we have seen that at lower initial energies that the QFI reaches first an intermediate plateau within the time $\tp$.  Here the value of $I^Q(t)$ is larger than the SNL but still scales like $N$ (see upper row of Fig.~\ref{fig:QFI_summary} and recall that the initial state is not entangled).  As the Wigner function in Fig.~\ref{fig:wigner_erg_LE} shows, the intermediate plateau corresponds indeed to the creation of a slightly squeezed state which rotates without spreading until the time $\ta$ is reached. Around $t=\ta$ the Wigner function suddenly spreads into a bimodal distribution. The latter does not isotropically cover the available region of phase space, which can be related to the mixed character of the underlying classical dynamics~\cite{altland2012equilibration}. One might suppose that this is also the reason why the characteristic timescales here do not seem to be scaling as the log of the system size, see upper right panel of Fig.~\ref{fig:timescales_N_all}.

\section{Conclusions}

Using the QFI, we studied the dynamics of multipartite entanglement in a fully-connected quantum many-body system across a regular-to-ergodic transition. The QFI allows to sharply distinguish the ergodic from the regular phase, as its asymptotic value, as well as the characteristic timescales, witness the transition both through their dependence on the control parameter $g$ and through their scaling with system's size $N$. 

The next set of investigations should involve the extension of the present analysis to many-body systems with finite-range interactions and also in the presence of disorder (where some results for QFI-dynamics in such systems have recently been discussed in the context of disordered ion chains~\cite{smith2016many}) and also in contact with external baths.

\begin{acknowledgments}

We are grateful to David Luitz for useful discussions and comments. Simulations were performed using the open source QuantumOptics.jl framework in Julia~\cite{kramer2018quantumoptics}; K.G. is grateful to David Plankensteiner for related discussions. K.G.\ acknowledges financial support from the National Science Centre Poland (NCN) under the ETIUDA scholarship (2017/24/T/ST2/00161).

\end{acknowledgments}

\end{document}